\documentclass[CEJP,DVI]{cej} 
\usepackage{amsmath}
\usepackage{textcomp}
\usepackage{hyperref}

\usepackage[dvips]{graphicx}
\usepackage{subfigure}
\usepackage[sort&compress]{natbib}

\title{Resonance States in an effective Chiral Hadronic Model}

\articletype{Research Article}

\author{P.~Rau\inst{1,}\inst{2}\email{rau@th.physik.uni-frankfurt.de},%
  J.~Steinheimer\inst{2},%
  S.~Schramm\inst{1,}\inst{2},%
  H.~St\"ocker\inst{1,}\inst{3}}

\institute{%
  \inst{1} Institut f\"ur Theoretische Physik, Goethe-Universit\"at,
  Max-von-Laue-Str.\ 1, 60438 Frankfurt am Main, Germany%
  \inst{2} Frankfurt Institute for Advanced Studies (FIAS),
  Ruth-Moufang-Str.\ 1, 60438 Frankfurt am Main, Germany%
  \inst{3} GSI Helmholtzzentrum f\"ur Schwerionenforschung GmbH,
  Planckstr.\ 1, 64291 Darmstadt, Germany%
}

\abstract{With an effective chiral flavour SU(3) model we show the
  effect of hadronic resonances on the QCD phase diagram. We state
  that varying the resonance couplings to the scalar and vector fields
  affects the order and location of the phase transition, the possible
  existence of a critical end point (CEP), and the thermodynamic
  properties. We present (strange) quark number susceptibilities at
  zero baryochemical potential and at three different points at the
  phase transition. Comparing results to lattice QCD, we state that
  reasonable large vector couplings limit the phase transition to a
  smooth crossover ruling out a CEP.}

\keywords{Effective Model \*\ Hadronic resonances \*\ QCD phase
  diagram \*\ Susceptibilities}
\pacs{12.38.-t, 11.30.Rd, 14.20.Gk}

\begin{document}
\maketitle
%
\section{Introduction}
\label{sec:introduction}
Both on the theoretical and on the experimental side there are great
efforts to study the largely unknown properties of the QCD phase
diagram with special attention to the phase transitions of strongly
interacting matter.  Most interesting are the transitions that restore
chiral symmetry and that generate deconfinement from hadrons to a
Quark Gluon Plasma (QGP) at high temperatures and
densities. Experimentally different regions of the phase diagram are
studied in ultra-relativistic heavy ion collisions at different beam
energies. Since QCD cannot be solved analytically, there are mainly
two groups of theoretical approaches to QCD. On the one hand there are
lattice QCD models, which are mostly limited to vanishing
baryochemical potentials, and on the other there are effective models
describing specific properties of QCD matter. Here, we present an
effective chiral flavour SU(3) model with hadronic degrees of freedom
including heavy resonance states and study the properties of the phase
diagram compared to lattice QCD results.
\section{Model}
\label{sec:model}
\begin{figure}[t]
  \centering 
  \subfigure{\includegraphics[width=.48\columnwidth]{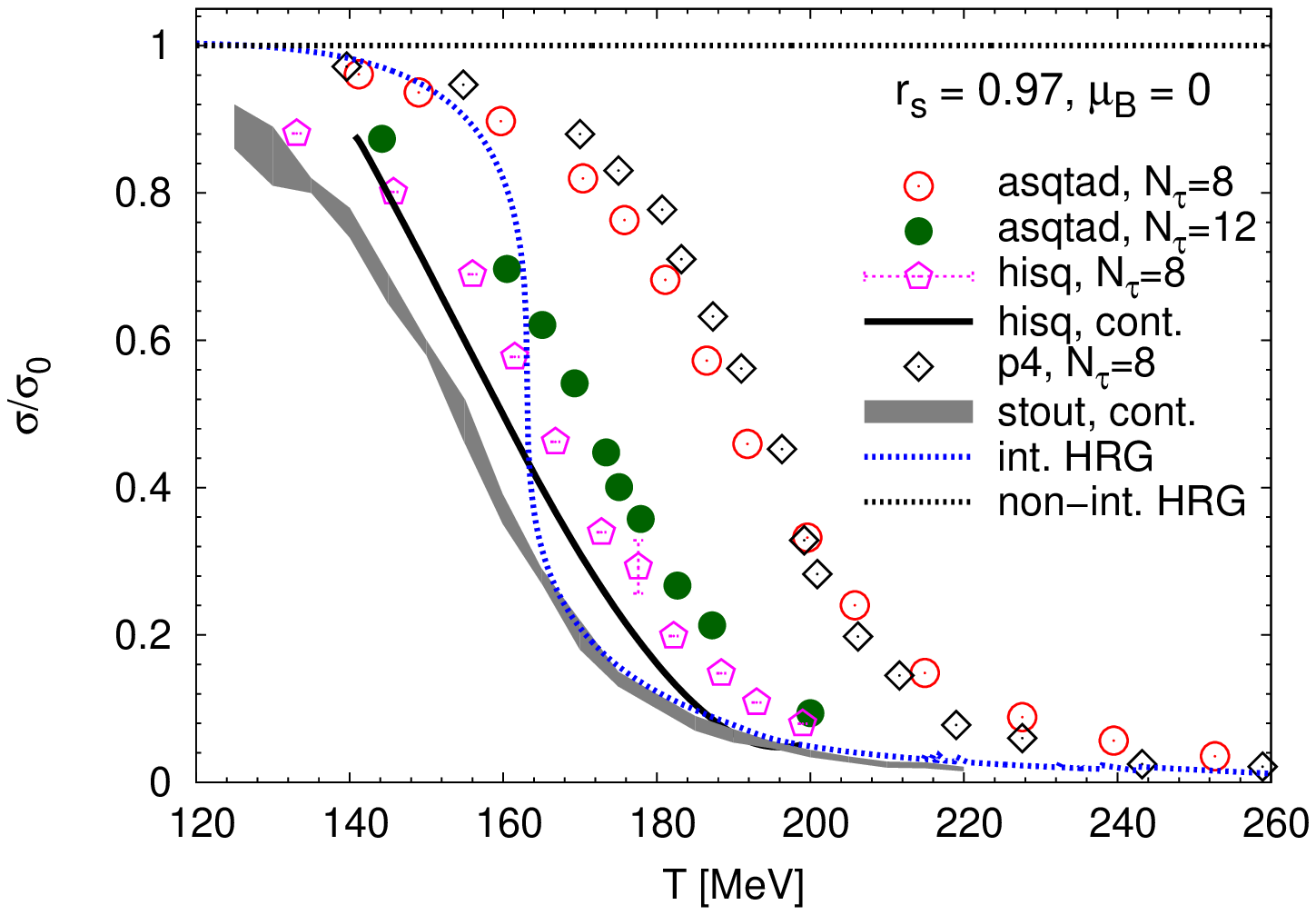}}
  \hfill
  \subfigure{\includegraphics[width=.48\columnwidth]{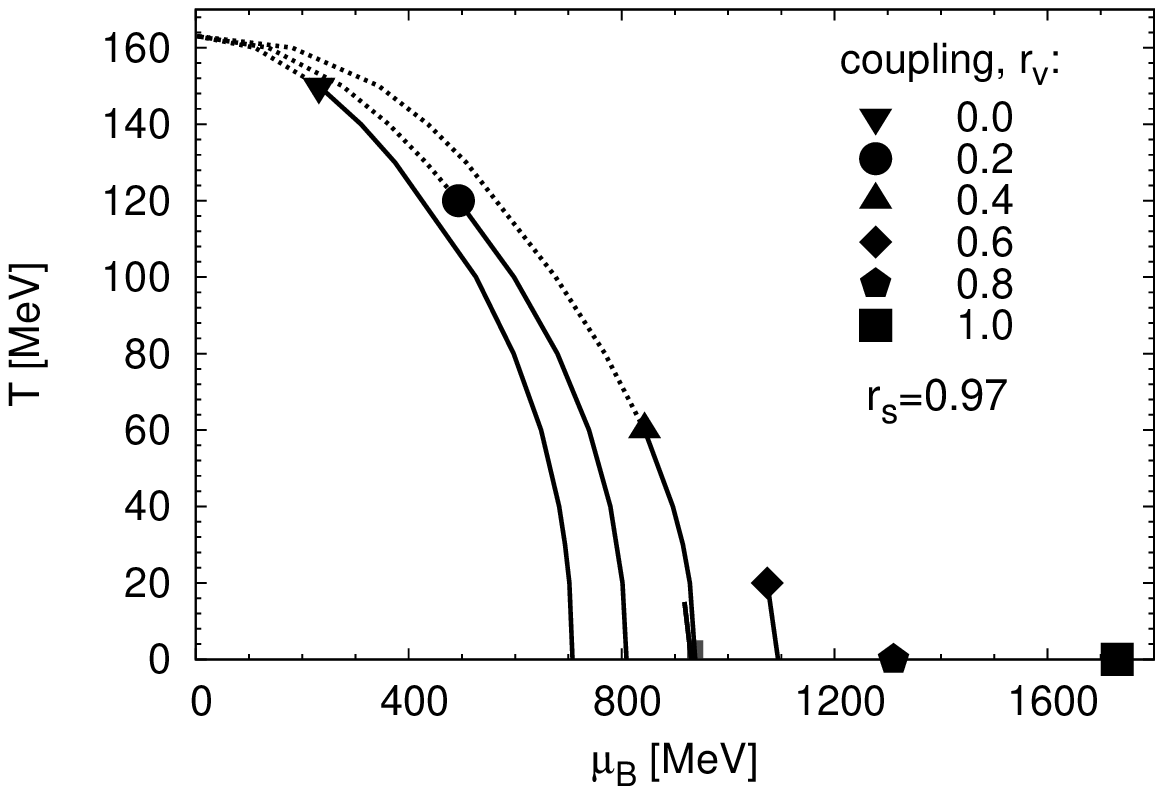}}
  \caption{(Color online) Left: Normalised chiral condensate order
    parameter together with lattice data for different actions. Right:
    Phase transition lines for different $r_v$. Dashed lines depict a
    crossover, solid lines a first order phase transition.}
  \label{fig:sigma_pt-lines}
\end{figure}
Our $SU(3)$-flavour $\sigma$-$\omega$-model with a non-linear
realization of chiral symmetry~\cite{model} is based on the mean field
Lagrangian $ \mathcal{L} = \mathcal{L}_{\rm kin} + \mathcal{L}_{\rm
  int} + \mathcal{L}_{\rm mes}$. With the hadrons' kinetic energy in
the first term, the interaction of baryons with scalar $\sigma$,
$\zeta$ and vector mesons $\omega$, $\phi$ is described by
\begin{equation}
  \label{eq:L_int}
  \mathcal{L}_{int} = -\sum_i \bar{\psi_i} \left( m^*_i + g_{i\omega}
    \gamma_0 \omega^0 + g_{i\phi} \gamma_0 \phi^0 \right) \psi_i .
\end{equation}
The index $i$ includes the baryon octet, decuplet, and heavier
resonances with masses $m \le 2.6$~GeV with a minimum three star
rating in the Particle Data Book. The term $\mathcal{L}_{\rm mes} =
\mathcal{L}_{\rm vec} +\mathcal{L}_{0} + \mathcal{L}_{\rm ESB}$
includes the vector and the scalar mesons' self interactions and the
explicit chiral symmetry breaking. The effective baryon masses
$m_{i}^* = g_{i\sigma}\sigma + g_{i\zeta}\zeta + \delta m_i$ are
generated by the coupling of baryons to the scalar meson fields
$\sigma$, $\zeta$ and by an explicit mass $\delta m_i$. At high
temperatures and baryonic densities this formalism leads to smaller
baryon masses and thereby to the restoration of chiral symmetry due to
the decreasing $\sigma$-field. The couplings of the baryon octet to
the mesonic fields and the mesonic potential are fixed to reproduce
the tabulated vacuum masses and well-known nuclear ground state
properties. The coupling strengths of the baryonic resonances are set
proportional to the nucleon couplings $g_{N}$ via the parameters
$r_{s,v}$ according to $g_{B \sigma, \omega} = r_{s,v} \cdot g_{N
  \sigma, \omega}$ and $g_{B \zeta, \phi} = r_{s,v} \cdot g_{N \zeta,
  \phi}$.  For simplicity and for generating the effective baryon
masses dynamically by the scalar fields, we keep the scalar coupling
fixed at $r_s = 0.97$. Thereby, we are able to reproduce the
particles' vacuum masses, except for the explicit mass $\delta m_i$,
and to ensure a smooth crossover transition at zero baryochemical
potential $\mu_B = 0$. However, the vector coupling parameter $r_v$,
which controls the abundances of the baryonic resonances at $\mu_B \ne
0$ , is varied in order to study its impact on the calculated phase
diagram and the thermodynamic properties. In contrast to this
interacting hadron resonance gas (HRG), we also perform calculations
for an ideal non-interacting HRG. In this particular case $r_{s,v}$
are set to zero and the masses of all particles are fixed at their
tabulated vacuum expectation value.\par
The grand canonical potential $\Omega / V = -\mathcal{L}_{\rm int} -
\mathcal{L}_{\rm meson} + \Omega_{\rm th}$ includes the thermal
contribution of all hadrons (i.e.\ all particles and antiparticles) in
the model. The effective baryochemical potential is defined as
$\mu^*_i = \mu_i - g_{i \omega} \omega - g_{i \phi} \phi$. From the
grand canonical potential all thermodynamic quantities follow, i.e.\
the pressure $p$, the energy and entropy densities $e$, $s$, and the
particle densities $\rho_i$.
\section{Results}
\label{sec:results}
\begin{figure}[t]
  \centering 
  \subfigure{\includegraphics[width=.48\columnwidth]{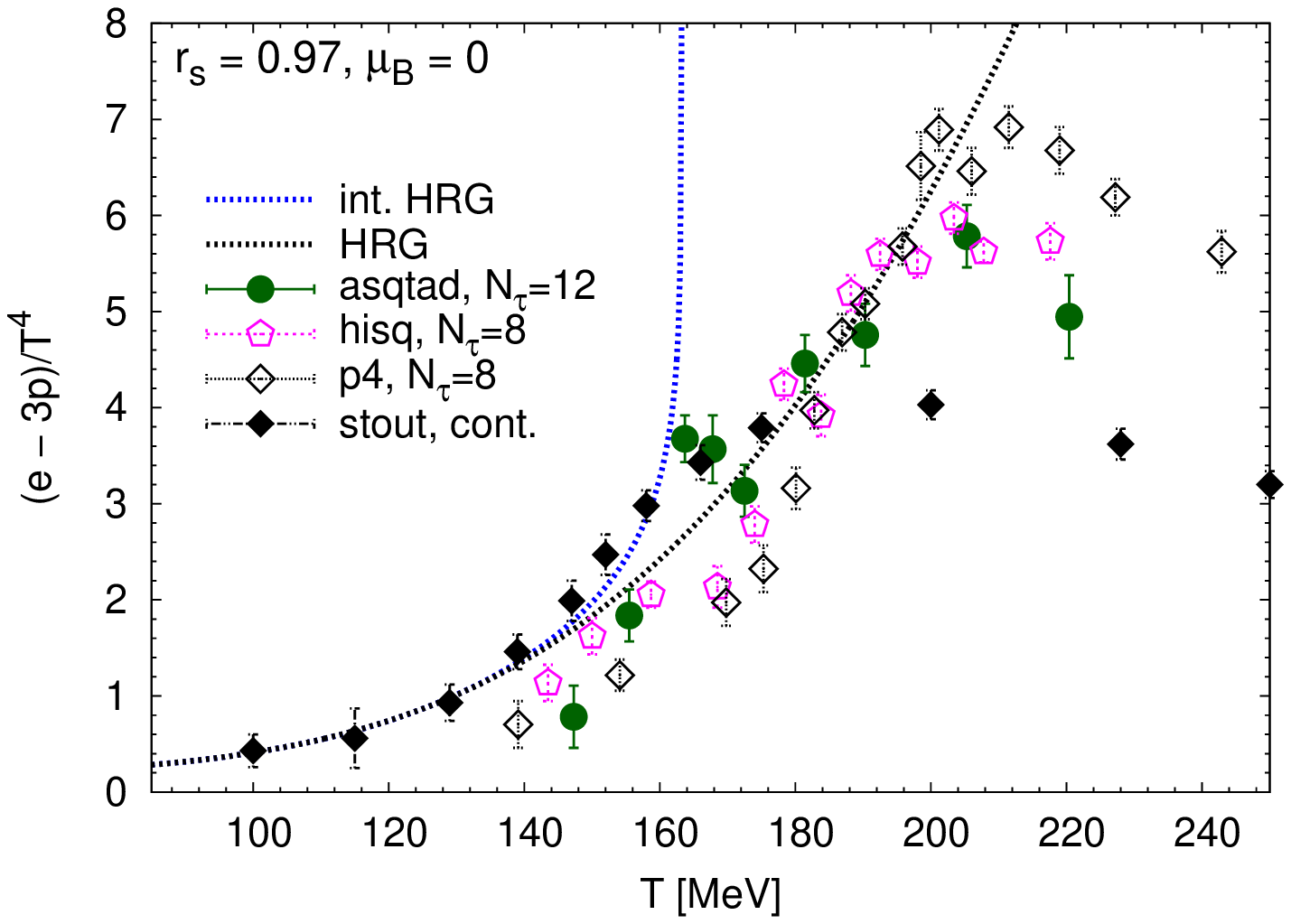}}
  \hfill
  \subfigure{\includegraphics[width=.48\columnwidth]{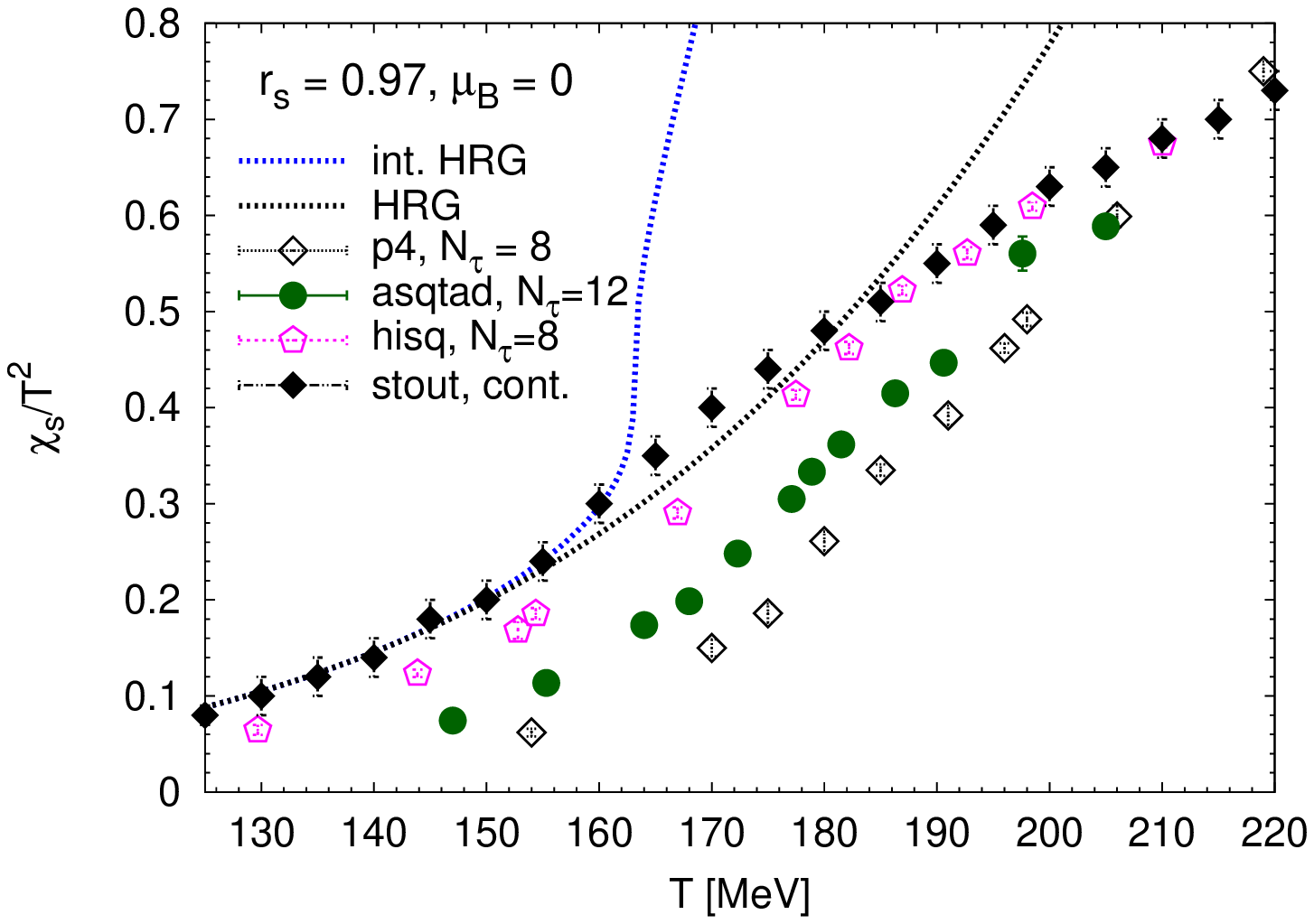}}
  \caption{(Color online) Interaction measure $(e-3\,p)/T^4$ (left)
    and the strange susceptibility $\chi_s/T^2$ (right) for the
    interacting (blue lines) and non-interacting (black lines) HRG
    compared to lattice data.}
  \label{fig:thermodyn_int_measure}
\end{figure}
The normalised chiral order parameter $\sigma/\sigma_0(T)$ from our
model is shown in Fig.~\ref{fig:sigma_pt-lines} (left) together with
lattice data from different actions and lattice
spacings~\cite{lattice}. With $r_s$ fixed, $\sigma/\sigma_0$ of the
interacting HRG (blue dotted line) exhibits a smooth crossover and a
critical temperature $T_c = 164$~MeV being in line with recent lattice
data for the continuum extrapolated HISQ and stout actions. Analysing
$\sigma/\sigma_0$ in the whole $T-\mu$-plane we get the phase
transition lines shown in Fig.~\ref{fig:sigma_pt-lines} (right) for
different values of $r_v$. Here, dashed lines depict crossover
transitions, dots a CEP, and full lines first order phase
transitions. We find that for reasonable values of $r_v \ge 0.4$, for
which the nuclear ground state is correctly located in the chirally
broken phase, the CEP moves to low temperatures or vanishes completely
($r_v > 0.6$) in favour of a broad crossover transition.\par
\begin{figure}[t]
  \centering 
\subfigure{\includegraphics[width=.48\columnwidth]{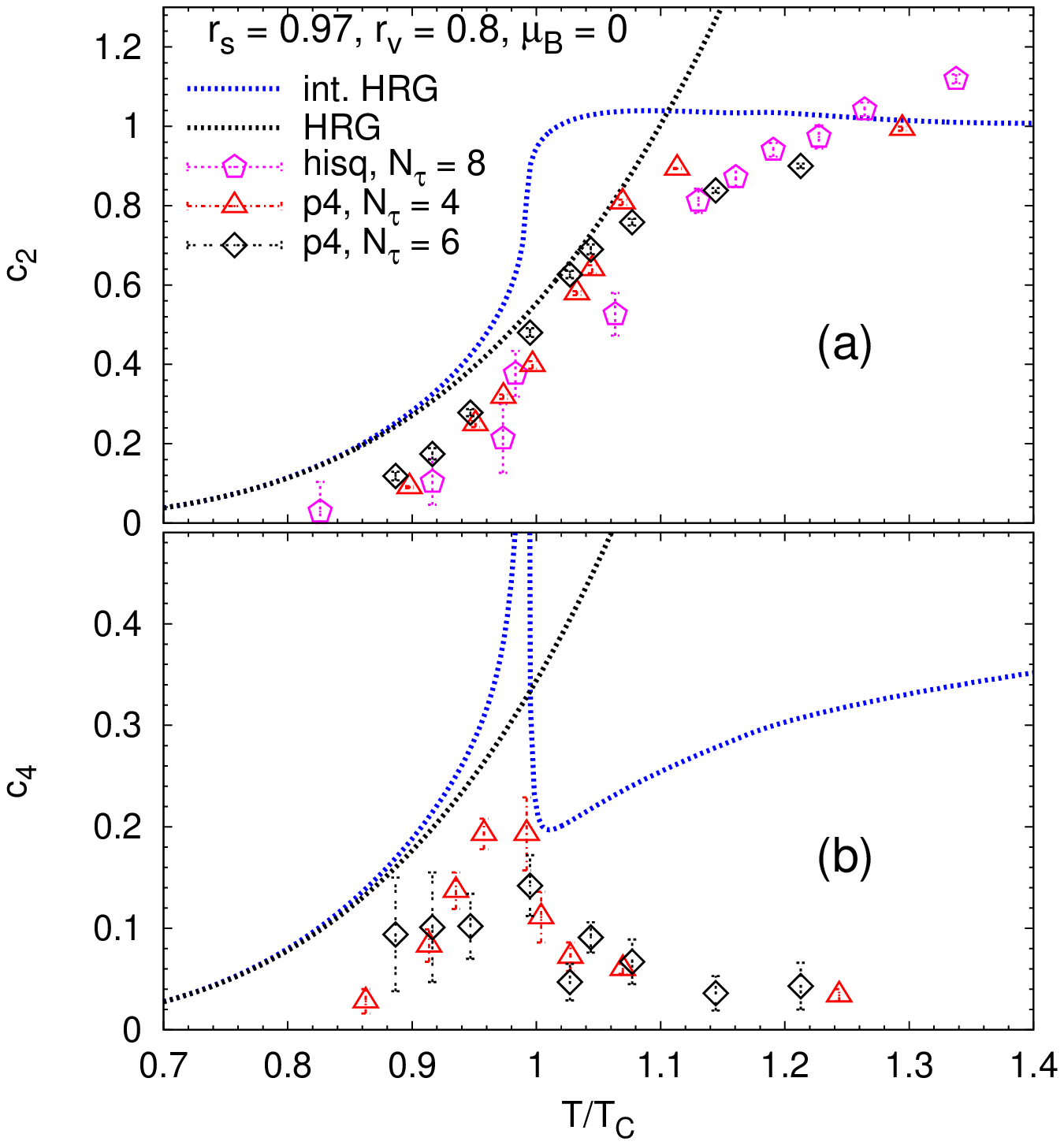}}
\hfill
\subfigure{\includegraphics[width=.48\columnwidth]{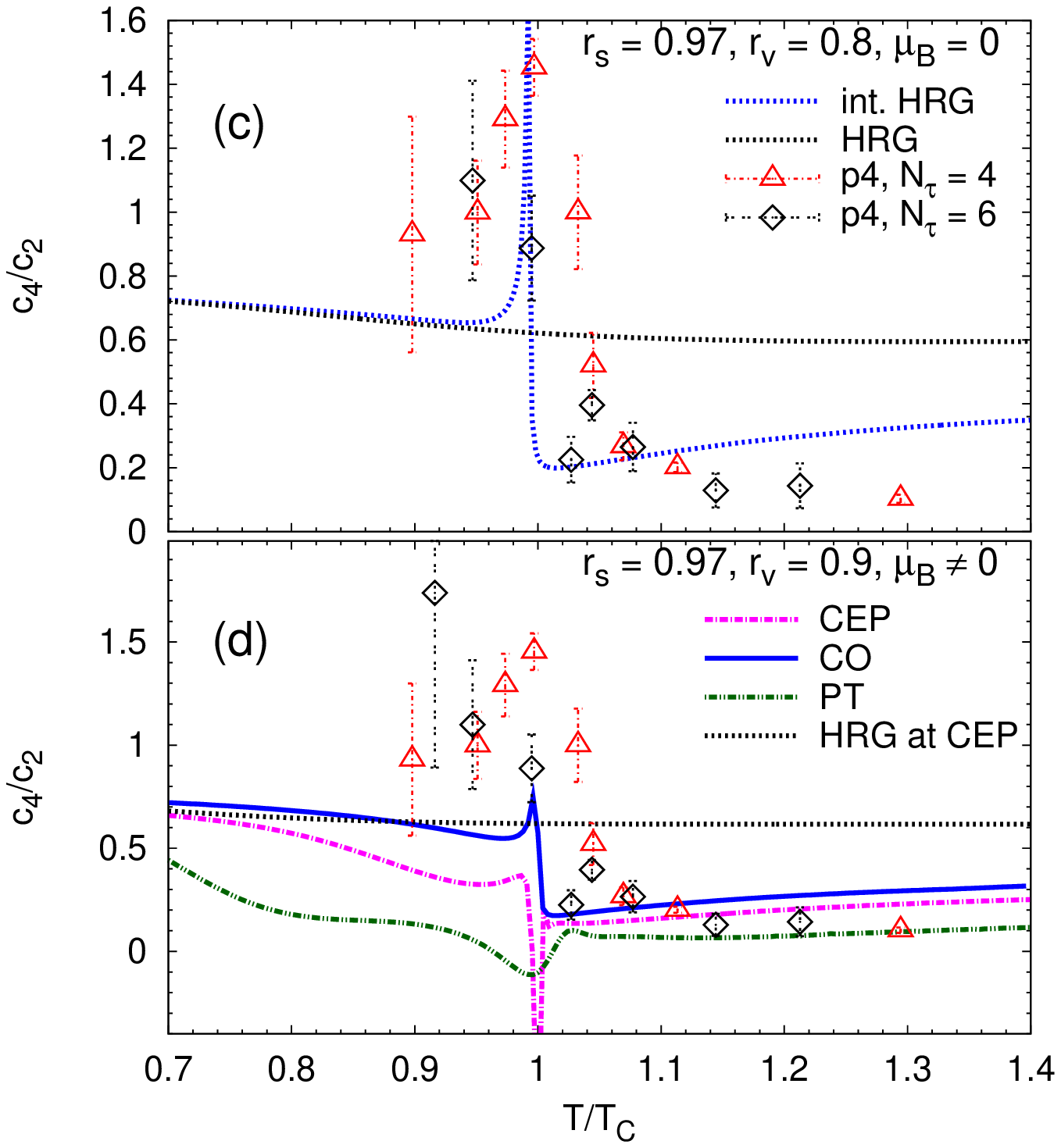}}
\caption{(Color online) Susceptibility coefficients $c_2$, $c_4$ ((a)
  and (b)) and the ratio $c_4/c_2$ as a function of $T/T_c$ at $\mu_B
  = 0$ (c) for the interacting (blue lines) and non-interacting (black
  lines) HRG for $r_v=0.8$ and at different points (CEP - $\mu_B =
  216$~MeV; CO - $\mu_B = 30$~MeV; PT - $\mu_B = 490$~MeV) on the
  phase transition for $r_v=0.9$ with lattice data (d).}
  \label{fig:susceps_ratio}
\end{figure}
Figure~\ref{fig:thermodyn_int_measure} shows the interaction measure
$(e-3\,p)/T^4$ (left) and the strange susceptibility $\chi_s/T^2$
(right) reflecting strange quark fluctuations~\cite{fluctuations} both
as functions of $T$. As expected, compared to the non-interacting HRG
the interacting HRG (blue lines) shows a much faster increase of both
quantities at $T_c$ due to more degrees of freedom. Both quantities
from the interacting HRG are in qualitatively good agreement with the
stout continuum limit.\par
Studying the second and fourth order quark number susceptibility
coefficients $c_2$, $c_4$ at $\mu_B = 0$ (Fig.~\ref{fig:susceps_ratio}
(a) and (b)), we find that fluctuations at $T_c$ get massively
suppressed by large couplings to the repulsive vector field
$r_v$. This is also reflected in the susceptibility ratio $c_4/c_2$
((Fig.~\ref{fig:susceps_ratio} (c)) which only comes close to lattice
data for reasonably high values of $r_v \ge 0.8$. This again supports
our previously mentioned preference for larger values $r_v$ and thus
limits the phase transition to a broad crossover in the whole
$T-\mu$-range (cf.\ Fig.~\ref{fig:sigma_pt-lines} (right)). For the
quark number susceptibilities we think, a comparison to not yet
available lattice data with continuum extrapolated stout action would
be very promising. A strong suppression of the susceptibility
coefficients is also seen at $\mu_B \ne 0$. When calculating the ratio
$c_4/c_2$ crossing the phase transition at different values of $\mu_B$
(Fig.~\ref{fig:susceps_ratio} (d)), we find that fluctuations are much
smaller compared to $\mu_B = 0$ which is again due to the repulsive
vector interactions.

\end{document}